\documentclass{aa}
\usepackage{graphicx}
\begin{document}

\title{Disk-Jet Connection in Cygnus X-3}
\author{M. Choudhury\inst{1}, A. R. Rao\inst{1}, S. V. Vadawale\inst{1}, C. H. Ishwara-Chandra\inst{2} \and A. K. Jain\inst{2}}
\offprints{M. Choudhury, \email{manojulu@tifr.res.in}}
\institute{Tata Institute of Fundamental Research, Mumbai-400005, India
	\and ISRO Satellite Center, Bangalore-560017, India}
\date{Received 19-11-2001/ Accepted 16-1-2001}
\authorrunning{Choudhury et al.}
\titlerunning{Disk-Jet Connection in Cygnus X-3}
\abstract{
We present the results of a detailed correlation study between 
the soft X-ray,  hard X-ray, 
 and radio emission (obtained from  $RXTE \ ASM$,  $BATSE$, and GBI
observations, respectively) of the
bright radio emitting Galactic X-ray binary Cygnus X-3.
We detect a very strong positive correlation between  the
soft X-ray and radio emission during the low-hard and
minor flaring periods of the source, and an anti-correlation between the
soft and hard X-ray emissions. We present statistical arguments to suggest
that the anti-correlation between the
radio and hard X-ray emission, reported earlier, is primarily due to
their correlation and anti-correlation, respectively, with the soft X-ray emission.
We make a wide band X-ray spectral study using the pointed RXTE
observations and detect a pivotal behaviour in the X-ray spectrum.
We argue that this X-ray
spectral pivoting is responsible for the anti-correlation between the
soft and hard X-ray emissions. The strong correlation between the 
soft X-ray and radio emission
suggests a close link between the accreting mechanism, 
plasma cloud surrounding the compact object and the radio emission. 
}

\maketitle

\section {Introduction}
Cygnus X-3 is the brightest radio source ever associated with an X-ray binary, 
in both quiescent and flare states (Waltman et al. \cite{wal95}). 
It is located at a distance of 9 kpc (Predehl et al. \cite{pre00}) in the 
Galactic plane in one of the arms. It exhibits multiple radio 
outbursts correlated with the high-soft state in the X-ray emission 
(Watanabe et al. \cite{wat94}). 
The VLBI observations of one such major flare revealed a core jet 
(Mioduszewski et al. \cite{mio01}), although it is not unambiguously 
resolved whether the jet is one-sided or two-sided (Marti et al. \cite{mar01}).
Watanabe et al. (\cite{wat94}) also hint at a correlation between the soft 
X-ray (as observed by ASM aboard the Ginga observatory) and the quiescent 
radio emission (as seen by the GBI), during the low-hard state. 
McCollough et al. (\cite{mcc99}) give a detailed correlation test between 
the hard X-ray (as observed by the BATSE aboard the CGRO) and the radio (GBI), 
in the various states of radio and X-ray emission. 
They report 1) anti-correlation between the radio and hard X-ray emission during 
the quiescent period, 2) correlation between the radio and hard X-ray emission 
during the major flaring period, and 
3) no correlation between the radio and hard X-ray emission during the minor flaring period. 

The (quasi) simultaneous observations of X-ray binaries 
in the radio and X-ray bands has led to the notion that 
the presence of radio jets is $ubiquitous$ in 
sources with black holes or low magnetic field ($\leq 10^9$G) neutron stars 
as compact objects and these sources 
show a definite connection between the accretion (inflow) 
mechanism and the jet (outflow) formation (see Fender et al. \cite{fen01}). 
Correlation between radio and soft X-ray emission has been established for at least 
two Galactic blackhole candidates (BHCs), Cygnus X-1 
(Brocksopp et al. \cite{bro99}) and GX 339-4 (Corbel et al. \cite{cor00}).  
In this $Letter$ we report the detection of a very strong correlation between 
the soft X-ray (as seen by the ASM aboard the RXTE) and the radio (GBI) 
during the low-hard and minor flaring state of Cygnus X-3, 
along with an  anti-correlation between 
the hard X-ray (BATSE) and both soft X-ray (ASM) and radio (GBI) emissions. 
The soft X-ray:radio correlation is definitely 
stronger and more significant than the anti-correlation between the hard X-ray 
and the soft X-ray (and radio). We also carry out a detailed wide
band spectral analysis, with available pointed RXTE observations
during the low-hard state. 

\begin{figure}[]
\resizebox{\hsize}{!}{\includegraphics[]{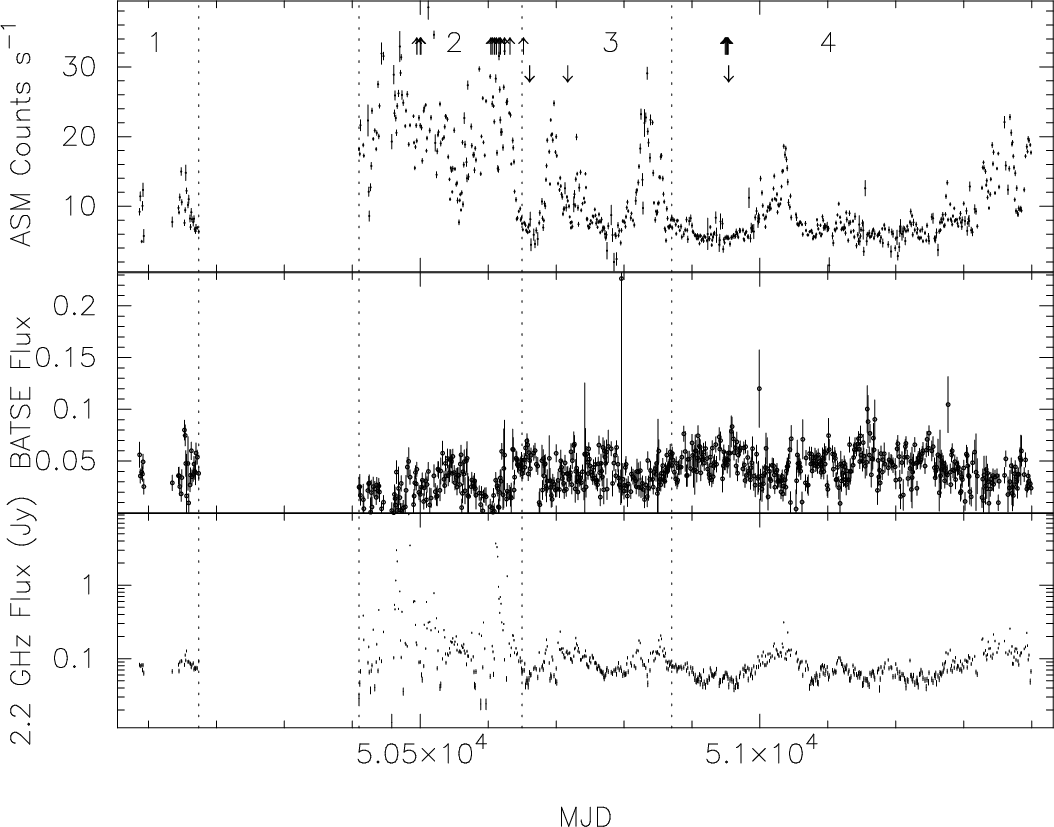}}
\caption{The combined simultaneous lightcurve of Cygnus X-3 
in the soft X-ray (ASM, top panel), hard X-ray (BATSE, middle panel) 
and the radio (GBI, bottom panel). The various `states' of the source
are separated by vertical dashed lines and identified with numbers.
The arrows on the  top panel give the start time of RXTE
pointed observations  and 
the three inverted arrows give the days for which wideband 
X-ray spectral studies are carried out.}
\label{fig1}
\end{figure}

\section {Data and Analysis}
The daily monitoring data are obtained from the archives of the 
respective observatories: soft X-ray, in the  energy range 2 -- 10 keV, from ASM 
(aboard RXTE), hard X-ray in the energy range 20 -- 600 keV from BATSE (aboard CGRO)
and radio at 2.2  and 8.3 GHz from GBI (Green Bank Interferometer, at Green Bank 
site, West Virginia, operated by NRAO). The pointed observations of both 
the narrow field  of view instruments aboard the RXTE, viz. PCA and HEXTE,  
are analysed for wide-band spectral study by combining the data from both, 
which include 129 channel PHA data from PCA standard-2 data (all PCUs added) 
and 64 channel data from the Cluster-0 of HEXTE. 
After analysing data from the Crab nebula, Vadawale et al. (\cite{vad01}) 
suggest using only Cluster-0 data from HEXTE and adding a systematic 
error of 2\% to PCA data for a proper fit, and we have followed the same recipe. 
The basic data reduction and analysis was carried out using FTOOLS (V5.0) and XSPEC (V11.0).

In Figure \ref{fig1} we plot the daily averaged lightcurve of Cygnus X-3 
as seen by ASM (top panel), BATSE (middle panel) and GBI (2.2 GHz, bottom panel), 
during the period when all these three detectors were simultaneously monitoring the source. 
Historically, the behaviour of radio emission in Cygnus X-3 is 
classified into: 1) quiescent period ($\sim$ 50--100 mJy), 
2) major flaring ($\geq$ 1Jy) with a preceding quenched state 
($\sim$ 10--20 mJy), and 3) minor flaring ($\geq$ 100--150mJy) 
with partial quenching state (see Waltman et al. \cite{wal95}; McCollough et al. \cite{mcc99}).
 Accordingly we have demarcated four regions in Figure 
\ref{fig1}, region 1 and 4 corresponding to the quiescent state
(although region 4 contains two minor flares along with the 
long quiescent period), region 2 corresponding to the major
flaring state and region 3 corresponding to the minor flaring
state.

\begin{figure}[t]
\resizebox{\hsize}{!}{\includegraphics[]{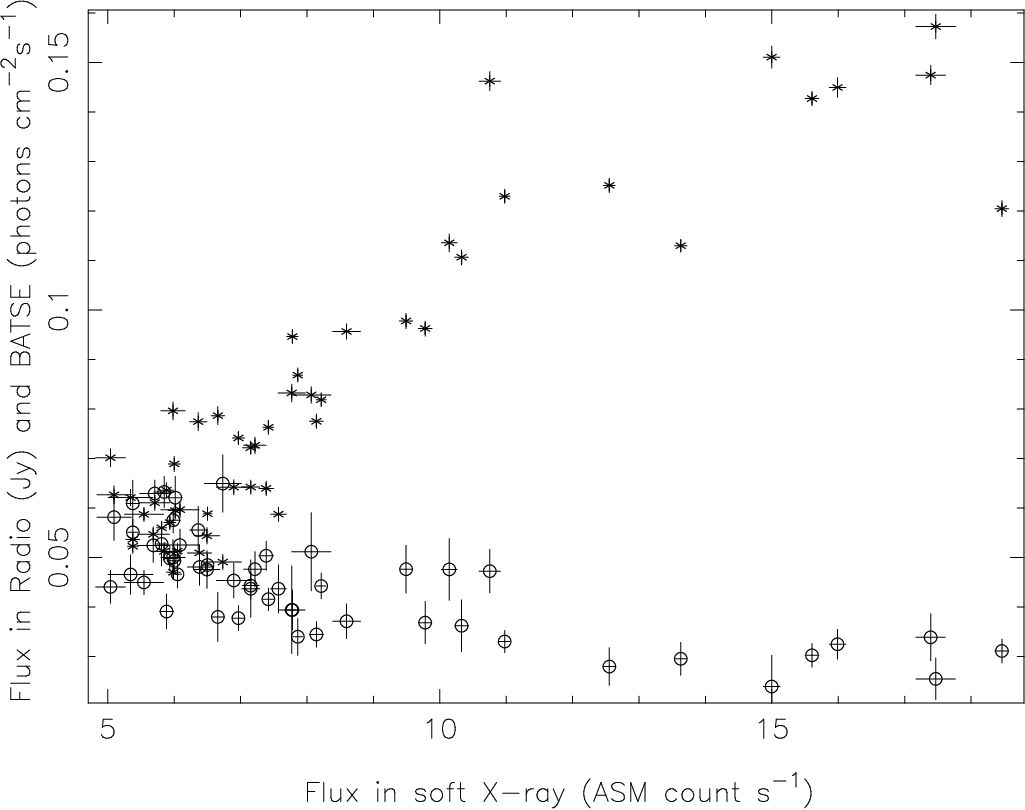}}
\caption{The variation of flux in radio (GBI, 2.2 GHz, shown as plus
signs) 
and hard X-ray (BATSE, 20-600 keV, shown as circles) 
with soft X-ray (ASM, 2-10 keV), during the low-hard state of Cygnus X-3 
(region 4 of Figure \ref{fig1}). Each data point is an average value
for 10 days.}
\label{fig2}
\end{figure}

To test for the correlation among the soft X-ray, hard X-ray and the radio emission we have used the Spearman Rank Correlation
(SRC) coefficient adapting the method prescribed by Macklin
(\cite{mac82}) and have derived 
the corresponding D-parameter, which gives the confidence level,
in terms of standard deviations, that the derived correlation is not due to the influence of the third parameter. 
Table 1 shows the SRC coefficient, null hypothesis probability
and the D-parameter, using 10 day averages of the data for the
different periods 
(and their combination) as demarcated in Figure \ref{fig1}. 
Reducing the number of days for averaging does not significantly change the results. 
The number of data points in region 1 (of Figure \ref{fig1}) is meagre and 
hence is not included in the correlation tests.

The most interesting result is 
that the soft X-ray and radio are very strongly correlated, 
with a very high significance level, in region 4 and regions 3 \& 4 combined. 
It can also be ruled out (at $>$ 5 $\sigma$ level) that this correlation
is influenced by the third parameter, the hard X-ray emission. Though
these two parameters are correlated even in the flaring state (region 2) at
a much reduced significance level, the correlation tests in this
region could be influenced by the high variability at time scales 
shorter than the period chosen for taking the averages (10 days). Hence
we concentrate on the results obtained for region 4 and region 3 \& 4
combined. 
It is also found that the soft X-ray is anti-correlated with the
hard X-ray emission. The anti-correlation between the radio and
hard X-ray emission, though strong,  could be influenced by the other two 
correlations, particularly when we examine the data for the region 
3 \& 4 combined.
The similarity of the behaviour of region 4 and regions 3 \& 4 combined 
suggest that the emission mechanism during the minor flaring and the quiescent 
period are the same, and hence these two can be clubbed together as one class.

\begin{figure}[]
\resizebox{\hsize}{!}{\includegraphics[]{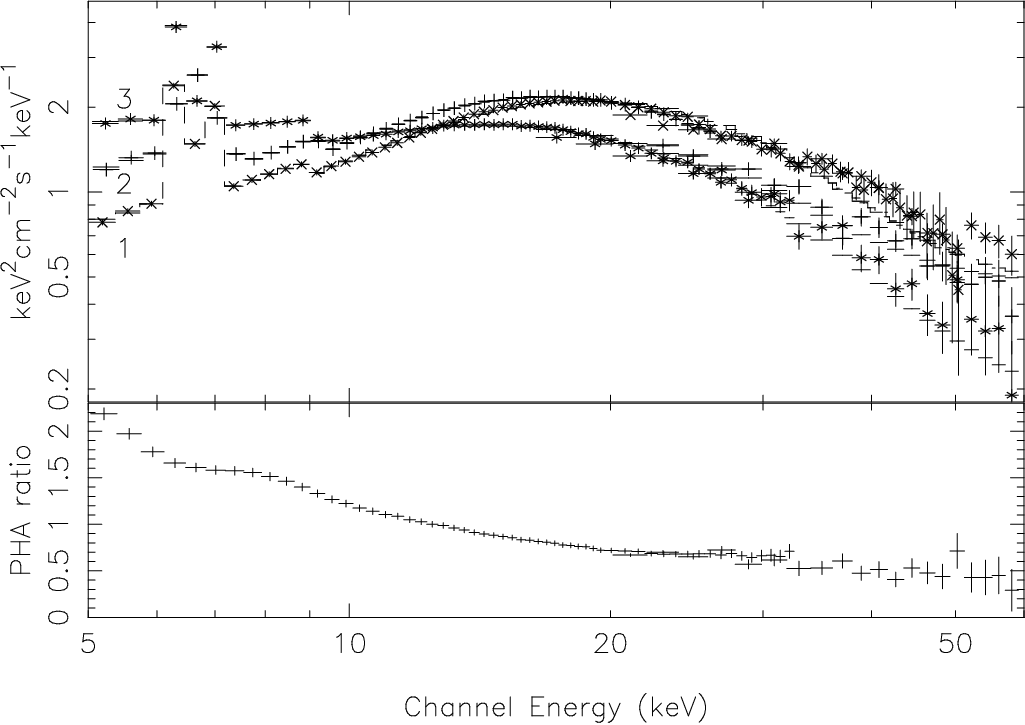}}
\caption{Top panel: The unfolded X-ray spectra of Cygnus X-3 during the 
radio quiescent period on three different occasions (1: MJD 50954; 2: MJD 50661; 3: MJD 50717).
Bottom panel: PHA ratio of spectra from the extreme observations.}
\label{fig3}
\end{figure}

Figure \ref{fig2}, showing the variation of flux in radio
(correlated) and hard X-ray (anti-correlated) with soft X-ray
flux during the low-hard state (region 4 of Figure \ref{fig1}), 
clearly demonstrates the simple monotonic dependence of the 
both with the soft X-ray (ASM) during the few hundred days. 
A total of 22 pointed observations of RXTE, of which 10 
are in the low-hard state, exist during this period of simultaneous monitoring (shown in the top panel of Figure \ref{fig1}). 
Only three groups 
of observations, separated by more than ten days, exist, in the low-hard state. We have selected 
representative observation from each group,
marked as inverted arrows in Figure \ref{fig1}, for a detailed
X-ray spectral analysis.
These three observations span the range of observed X-ray and
radio fluxes and some salient features of these observations 
are given in Table 2.
The three unfolded spectra are overlaid on the top panel of Figure \ref{fig3} with 
the PHA ratios for the two extreme spectra shown  in the bottom panel. 

During the low-hard state, encompassing the regions 3 \& 4 of Figure \ref{fig1}, the continuum spectra (5 -- 150 keV) is best described by
the Comptonization of seed photons from a thermal multi-coloured 
accretion disk by a thermal Comptonizing plasma cloud 
(CompST-Sunyaev \& Titarchuk \cite{sun80}; Nakamura et al. \cite{nak93}; Rajeev et al. \cite{raj94}) along with a non-thermal powerlaw emission (Choudhury \& Rao \cite{cho01}).
The resolution of the three Fe lines (Kitamoto et al. \cite{kit94}) 
and the two absorption edges (Rajeev et al. \cite{raj94}) are beyond the 
capability of the PCA, hence we fix the relative separation of line and 
edge energies as reported by Rajeev et al. (\cite{raj94}); Nakamura et al. (\cite{nak93}); and Kitamoto et al. (\cite{kit94}),
and treat the edge energy at 7.1 keV and the normalization of all the 
lines and edges as the variable parameters in the fit. 
The best fit values of the electron temperature (kT$_e$) of the
Compton cloud and the powerlaw photon index ($\Gamma_X$) 
for the three spectra are 
given in Table 2B. Other simple models like cutoff powerlaw (along with a powerlaw),
broken power-law, etc. do not consistently fit all the observed spectra with
physically feasible parameter values and give much inferior fits. 
It can be seen from the figure that there is a systematic change in the
shape of the spectrum with increasing soft X-ray flux (and radio flux, 
see Table 2). The X-ray spectrum, when the radio flux is low, is quite 
flat in the 20 -- 40 keV
region and the spectral curvature increases with the soft X-ray flux.

\begin{table}[t]
\caption{The Spearman Rank  Correlation (SRC) coefficient, null-hypothesis 
probability and D-parameter between the radio, soft X-ray and hard 
X-ray for different periods demarcated in Figure \ref{fig1}.}
\begin{tabular}{lccc}
\hline   \hline
 & SRC coeff. &  Null Prob. & D-Parameter \\
\hline
{\bf Region 4}      
                                         \\  \cline{1-4}
ASM:GBI & 0.84  & 6.3$\times10^{-11}$ & 5.2 \\
GBI:BATSE & $-$0.75 & 9.2$\times10^{-11}$ & $-$2.7 \\
ASM:BATSE & $-$0.74 & 3.1$\times10^{-11}$ & $-$2.2 \\
{\bf Region 3}      
                                       \\ \cline{1-4}
ASM:GBI & 0.66 & 8.3$\times10^{-4}$ & 2.7 \\
GBI:BATSE & $-$0.43 & 4.7$\times10^{-2}$ & 0.4 \\
ASM:BATSE & $-$0.71 & 1.9$\times10^{-4}$ & $-$3.2 \\
{\bf Region 2}      
                                       \\ \cline{1-4}
ASM:GBI & 0.56 & 3.5$\times10^{-3}$ & 4.1 \\
GBI:BATSE & 0.10 & 6.3$\times10^{-1}$ & 2.7 \\
ASM:BATSE & $-$0.50 & 1.1$\times10^{-2}$ & $-$3.8 \\
{\bf Region 3 \& 4}   
                                         \\  \cline{1-4}
ASM:GBI & 0.83 & 2.1$\times10^{-20}$ & 6.2 \\
GBI:BATSE & $-$0.72 & 4.7$\times10^{-13}$ & $-$1.5 \\
ASM:BATSE & $-$0.79 & 4.1$\times10^{-17}$ & $-$3.8 \\
\hline
\end{tabular}
\end{table}

\begin{table}
\caption{ The observed flux and X-ray spectral parameters of Cygnus X-3 during
the three pointed RXTE observations.}
\begin{tabular}{lccc}
\hline  \hline
\multicolumn{1}{l}{} & \multicolumn{3}{c}{MJD} \\
\cline{2-4} 
 & 50717 & 50661 & 50954 \\
\hline
{\bf A.}Flux & & & \\
\hline
ASM (cts s$^{-1}$) & 11.11 & 8.18 & 5.37 \\
BATSE (ph cm$^{-2} s^{-1}$) & 0.038 & 0.051 & 0.058 \\ 
GBI-2.2GHz (mJy) & 115 & 64 & 43 \\
GBI-8.3GHz (mJy) & 165 & 73 & 53 \\
\hline 
\multicolumn{4}{l}{{\bf B.} Best fit parameters for CompST+power law} \\
\hline 
kT$_e$ (keV) & 5.09$\pm0.38$ & 4.37$\pm0.04$ & 4.87$\pm0.08$ \\
$\Gamma_X$ & 2.55$\pm0.22$ & 2.19$\pm0.05$ & 2.01$\pm0.04$ \\
$\chi^2_\nu$(d.o.f.) & 0.74(86) & 1.80(59) & 1.42(108) \\
\hline 

\end{tabular}
\end{table}

\section{Discussion and Conclusion}

Although Galactic BHCs exhibit radio jets during the low-hard state,
viz. Cygnus X-1 (Brocksopp et al. \cite{bro99}), GX339-4
(Corbel et al. \cite{cor00}), GRS1915+105 (Dhawan et al.
\cite{dha00}), XTEJ1550-534 (Corbel et al. \cite{cor01}),
1E1740.7 (Mirabel et al. \cite{mir92}), GRS1758-258
(Rodriguez et al. \cite{rod92}), a detailed correlation among the radio, soft and hard X-ray emission is done for a very few
sources, viz. Cygnus X-1 and GX339-4, and their behaviour is
quite different from that of Cygnus X-3. Cygnus X-3 is the only
source that shows such a strong correlation between soft X-ray
and radio, and it is also the only source to distinctly show the
anti-correlation between the soft and hard X-rays, during the
low-hard state.
Watanabe et al. (\cite{wat94}) have examined the association of
the X-ray (from Ginga) and radio emission from Cygnus X-3 and 
though there was a hint of correlation between the soft X-ray and
radio flux in the hard state (see their Figure 6), a detailed
correlation was not presented, presumably because of the sparse
sampling of the source by the Ginga satellite.

 The quiescent state of Cygnus X-3 has persistent (flat spectrum) radio emission 
(60 -- 100 mJy). This state in Cygnus X-3 is likely to be similar to the ``plateau''radio state seen in the most active micro-quasar GRS 1915+105 which shows
flat spectrum radio emission for extended durations and the
radio emission is identified with a compact jet of size $\sim$10 AU (Dhawan et al. \cite{dha00}). 
The spectral changes in association with the radio emission is
also quite similar to GRS 1915+105 (Vadawale et al. \cite{vad01}). 

Recently it has been suggested that X-ray emission from 
BHCs GRS1915+105 (particularly during the ``plateau'' state) and 
XTE J1118+480 could be arising from synchrotron emission from the
base of the jet (Vadawale et al. \cite{vad01}; Markoff et al. \cite{mark01}). 
Therefore it
leads to a speculation that some of the X-ray flux in Cygnus X-3 also 
could be arising from synchrotron emission from the base of the jet. This 
can explain the correlation between soft X-ray and radio fluxes but fails 
to explain the anti-correlation between soft and hard X-ray fluxes.

It is also possible that the soft X-ray emission is due to the 
accretion disk (directly or indirectly)
and the observed correlation is due to a connection between the accretion 
disk and the jet emission.
The X-ray emission from Cygnus X-3 is highly obscured and the
bulk of the X-ray emission below 5 keV is due to the 
emission-line dominated photo-ionized plasma surrounding the
compact object (Paerels et al. \cite{pae00}; Kawashima \& Kitamoto \cite{kaw96}).
Hence there is no clear evidence for the
disk blackbody emission, commonly seen as an X-ray spectral component 
in the soft X-ray region in other BHCs. Our spectral analysis above 5 keV has
identified two spectral components and these, by analogy with other BHCs, can be 
identified with thermal/non-thermal Comptonization (Zdziarski et al. \cite{zdz01}; 
Gierlinski et al. \cite{gie99}) occurring in the source (or due to X-ray synchrotron emission - see above).
If we assume that the region of the Comptonization is confined to a small region
near the compact object, we can qualitatively  explain the observed correlations
under the Two Component Accretion Flow (TCAF) model of Chakrabarti (\cite{cha96}),
in which the Compton spectrum originates from a region close to
the compact object, confined within the Centrifugal Boundary Layer (CENBOL). 
At low accretion rate, the CENBOL is far away from the compact object, the
spectrum is harder with lower outflow (see Das and Chakrabarti \cite{das98}). On 
increasing the accretion rate the 
CENBOL comes closer to the compact object with greater outflow,
giving rise to increased radio emission.
Though this model qualitatively explains the observed correlations, we
must add here that the thermal Compton model is only an approximation and a correct
Comptonization model requires an accurate description of the geometry of the 
emission region.

\section*{Acknowledgements} This research has made use of data obtained through the HEASARC Online Service, provided by the NASA/GSFC, and the Green Bank Interferometer, a facility of the National Science Foundation operated by the NRAO in support of NASA High Energy Astrophysics Programs.
We thank J. S. Yadav for useful suggestions and H. Falcke, the referee of the
paper, whose critical comments helped in significantly improving the quality of
the work.
AKJ is grateful to P. S. Goel, Director, ISAC and K. Kasturirangan,
Chairman, ISRO, for their constant encouragement and support during the course of this
work.

\end{document}